\documentclass{iau_FM}
\usepackage{graphicx}

\title[Solar models with Los Alamos ATOMIC opacities]{Sound speed and oscillation frequencies for solar models evolved with Los Alamos ATOMIC opacities}

\author[Guzik et al.]   
{Joyce A. Guzik$^1$, C.J. Fontes$^1$, P. Walczak$^2$, S.R. Wood$^3$, K. Mussack$^1$, 
 \and E. Farag$^{1,4}$}

\affiliation{$^1$Los Alamos National Laboratory, Los Alamos, NM USA 87545 \\ email: {\tt joy@lanl.gov}
 \\[\affilskip]$^2$Instytut Astronomiczny, Uniwersytet Wroc{\l}awski, Wroc{\l}aw, Poland
  \\[\affilskip]$^3$University of Oregon, Eugene, OR  USA 97403
\\[\affilskip]$^4$Ohio State U., Columbus, OH  43210 USA
}

\pubyear{2015}
\jname{Astronomy in Focus, Volume 2} 
\editors{Piero Benvenuti, ed.}
\begin{document}

\maketitle

\begin{abstract}

Los Alamos National Laboratory has calculated a new generation of radiative opacities (OPLIB data using the ATOMIC code) for elements with atomic number Z=1-30 with improved physics input, updated atomic data, and finer temperature grid to replace the Los Alamos LEDCOP opacities released in the year 2000.
We calculate the evolution of standard solar models including these new opacities, and compare with models evolved using the Lawrence Livermore National Laboratory OPAL opacities (\cite{OPAL}). We use the solar abundance mixture of \cite{AGSS09}.
The Los Alamos ATOMIC opacities (\cite[Colgan \etal\ 2013a, 2013b, 2015]{colgan2013a, colgan2013b, colgan2015}) have steeper opacity derivatives than those of OPAL for temperatures and densities of the solar interior radiative zone.
We compare the calculated nonadiabatic solar oscillation frequencies and solar interior sound speed to observed frequencies and helioseismic inferences.
The calculated sound-speed profiles are similar for models evolved using either the updated Iben evolution code (see \cite{Guzik2010}), or the MESA evolution code (\cite{MESA}).
The LANL ATOMIC opacities partially mitigate the `solar abundance problem'.

\keywords{atomic data: opacities; Sun:  interior; Sun:  evolution; Sun:  oscillations}

\end{abstract}

\firstsection 
              
\section{New Los Alamos ATOMIC Opacities}

Los Alamos opacity library (OPLIB) tables have been generated using a new code called ATOMIC for the first 30 elements of the periodic table and are now available on-line\footnote{http://aphysics2.lanl.gov/opacity/lanl}.  This new OPLIB release includes improvements such as a more accurate equation-of-state treatment, refined temperature grid, and significant fine-structure detail in the atomic physics calculations (for details, see \cite{colgan2013a, colgan2013b, colgan2015}).  The opacities used here were generated by mixing the pure-element OPLIB tables under the assumption of electron-temperature and electron-degeneracy equilibrium.

\begin{figure}[b]
\begin{center}
 \includegraphics[width=3.4in]{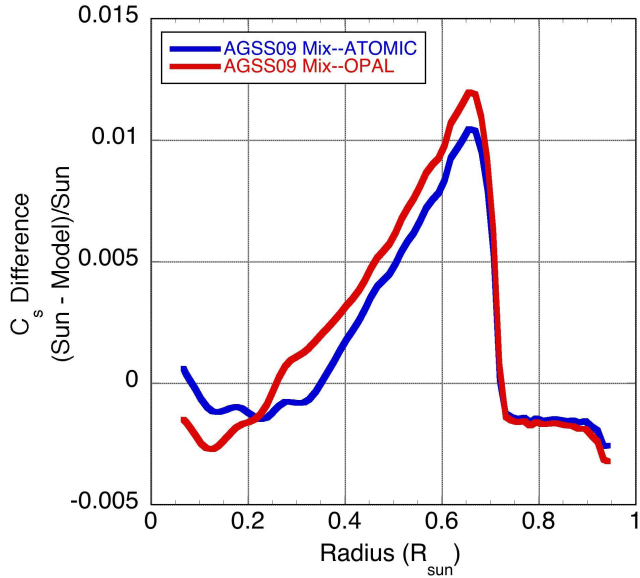} 
 \caption{Helioseismically inferred minus calculated sound speed  differences vs. radius for solar models using the ATOMIC and OPAL opacities with the AGSS09 abundance mixture.}
   \label{soundspeed}
\end{center}
\end{figure}

\begin{figure}[b]
\begin{center}
 \includegraphics[width=3.4in]{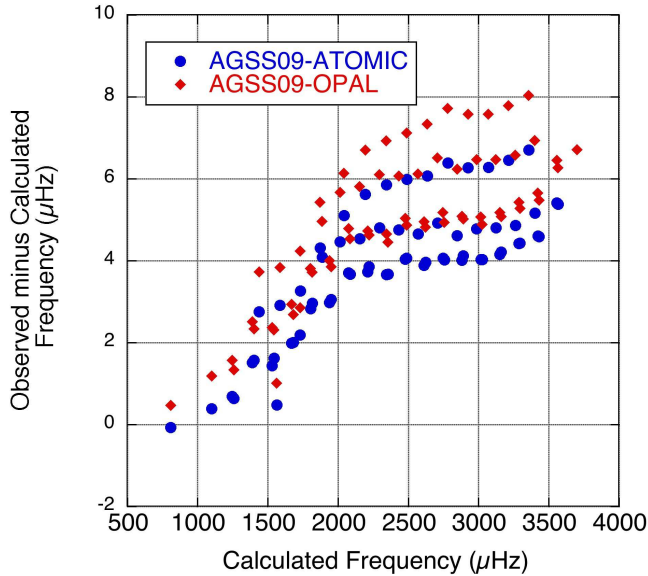} 
 \caption{Observed minus calculated nonadiabatic frequencies for solar oscillations of degrees $l$ = 0, 2, 10, and 20 using the OPAL or ATOMIC opacities with AGSS09 element abundance mixture.  Observations are from BiSON (\cite{Chaplin1998}), LowL (\cite{SchouTomczyk1996}) or GOLF (\cite{Garcia2001}). Frequencies are calculated using the \cite{Pesnell1990} code.}
   \label{O-C}
\end{center}
\end{figure}

\begin{figure}[b]
\begin{center}
 \includegraphics[width=3.4in]{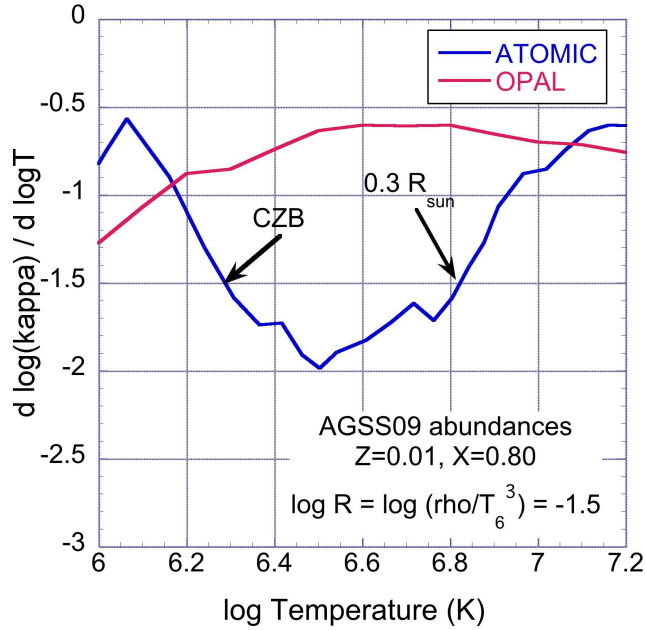} 
 \caption{Logarithmic temperature derivative of opacity for ATOMIC and OPAL opacities for solar radiative interior temperatures.  The larger opacity gradients at the convection zone base (log T = 6.3) and at $\sim$0.2-0.3 R$_\odot$ in the solar interior are responsible for the differences in sound speed gradient for calibrated solar models seen in Fig. \ref{soundspeed}.}
\label{gradient}
\end{center}
\end{figure}

\section{Solar Evolution and Oscillation Results}

We evolved solar models using both the Lawrence Livermore National Laboratory OPAL opacities (\cite{OPAL}) and the new Los Alamos ATOMIC opacities, for the AGSS09 (\cite{AGSS09}) solar abundance mixture.  We adjusted the initial helium mass fraction (Y), initial mass fraction of elements heavier than hydrogen and helium (Z), and mixing-length to pressure-scale-height ratio ($\alpha$) to calibrate the model to the observed solar luminosity and radius at the current solar age, as well as to the observed Z/X ratio of the AGSS09 abundances (see \cite{Guzik2010} for details of the modeling procedure).  Table \ref{tab1} lists the calibration parameters, final photospheric Y, Z abundances, and envelope convection-zone base radii.

\begin{table}
  \begin{center}
  \caption{Calibration parameters and properties of solar models evolved using OPAL or ATOMIC opacities, with the AGSS09 abundance mixture.}
  \label{tab1}
  \begin{tabular}{lcc}\hline 
 & {\bf OPAL} & {\bf ATOMIC} \\ 
 \hline
{\bf Y$_{initial}$}  & 0.2641 & 0.2570 \\
{\bf Z$_{initial}$}  & 0.0150 & 0.0151 \\
{\bf $\alpha$ }  & 2.0118 & 2.0637 \\
\\
{\bf Y$_{conv. zone}$$^a$}  & 0.2345 & 0.2283 \\
{\bf Z$_{conv. zone}$}  & 0.0135 & 0.0136 \\
{\bf R$_{conv. zone~base}$$^b$ (R$_{\odot}$)} & 0.7264 & 0.7251 \\
 \hline
  \end{tabular}
 \end{center}
 $^a$Helioseismically inferred convection-zone Y is 0.248 $\pm$ 0.003 (\cite{BasuAntia2004})\\
 $^b$Helioseismically inferred convection-zone radius is 0.713 $\pm$ 0.001 R$_{\odot}$ (\cite{BasuAntia2004})
\end{table}

Figure \ref{soundspeed} shows the helioseismically inferred (\cite{Basu2000}) minus calculated sound-speed difference vs. solar radius for two models evolved using the updated Iben evolution code and AGSS09 abundance mixture.  The sound-speed agreement is slightly better for the model using the new LANL ATOMIC opacities compared to that using the LLNL OPAL opacities.  We found similar results for models evolved with the MESA code (\cite{MESA}) using the same opacity tables and abundance mixture.  Figure \ref{O-C} shows the observed minus calculated vs. calculated low-degree nonadiabatic solar oscillation frequencies for the two models, also showing slightly improved agreement using the ATOMIC opacities compared to the OPAL opacities. 

For the same temperature, density, and abundance mixture, the ATOMIC opacities are actually lower in the radiative interior than the OPAL opacities, although they are higher than the OPAL opacities for conditions in the solar convective envelope that do not affect the solar structure since convection is transporting nearly all of the solar luminosity.  It turns out that the difference in opacity {\it derivatives} between OPAL and ATOMIC for solar interior conditions is responsible for the difference in solar structure. Figure \ref{gradient} shows that the opacity derivative with respect to temperature for the ATOMIC opacities is steeper at log T = 6.3 (CZ base) than that of the OPAL opacities, resulting in a slightly deeper convection zone and change in the sound-speed gradient at the convection-zone base.  In addition, the opacity derivative is steeper at log T= 6.8 (solar radius $\sim$0.2-0.3 R$_\odot$), also changing the sound-speed gradient at this location.  

\section{Conclusion and Future Work}

The solar models presented here evolved with the updated Iben code use the \cite{Ferguson2005} low-temperature opacities; we would like to use low-temperature opacities created for the AGSS09 mixture. We would like to investigate the effects of ATOMIC opacities on other types of pulsating stars, as has been done recently for B stars (Walczak et al. 2015). We would like to test implications of opacity changes based on ongoing experiments using laser and pulsed-power facilities (see, e.g., \cite{Bailey2015}).

\acknowledgments
The Los Alamos National Laboratory is operated by Los Alamos National Security, LLC for the National Nuclear Security Administration of the U.S. Department of Energy under Contract No. DE-AC52-06NA25396. P. Walczak was supported for this work under European Community's Seventh Framework Programme (FP7/2007-2013) under grant agreement No. 269194. We obtained LLNL opacities from the Lawrence Livermore National Laboratory OPAL Opacity Web site:  http://opalopacity.llnl.gov/opal.html

\end{document}